\documentstyle[12pt]{article}

\newcommand{\be}{\begin{equation}}
\newcommand{\ee}{\end{equation}}

\newcommand{\ba}{\begin{eqnarray}}
\newcommand{\ea}{\end{eqnarray}}

\newcommand{\la}{\lambda}
\newcommand{\al}{\alpha}

\newcommand{\r}{\rho}
\newcommand{\D}{\Delta}
\newcommand{\Tr}{\rm Tr}
\newcommand{\g}{\gamma}
\newcommand{\e}{\epsilon}

\begin{document}

\hsize36truepc\vsize51truepc
\hoffset=-.4truein\voffset=-0.5truein
\setlength{\textheight}{8.5 in}

\begin{titlepage}
\begin{center}
\hfill LPTENS-99/33\\

\vskip 0.6 in
{\large  Characteristic polynomials of random matrices}
\vskip .6 in
       {\bf Edouard Br\'ezin \footnote{$  ${\it
 Laboratoire de Physique Th\'eorique de l'\'Ecole Normale
Sup\'erieure, Unit\'e Mixte de Recherche 8549 du Centre National de la
Recherche
Scientifique et de l'\'Ecole Normale Sup\'erieure,
24 rue Lhomond, 75231 Paris Cedex 05, France.\\ {\bf brezin@physique.ens.fr}}}}
            {\it{and}}\hskip 0.3cm
       {\bf {Shinobu Hikami}}  \footnote{ $  ${\it{ Department of
Basic Sciences, University of Tokyo\\ Meguro-ku, Komaba 3-8-1, Tokyo
153,
Japan.\\ {\bf hikami@rishon.c.u-tokyo.ac.jp}}}}\\

\vskip 0.6 in

{\bf Abstract}
\end{center}
\vskip 14.5pt

{\leftskip=0.5truein\rightskip=0.5truein\noindent
{\small

Number theorists have  studied extensively the connections between the
distribution
of zeros of the Riemann $\zeta$-function, and  of some generalizations, with
the statistics of the
eigenvalues of large random matrices. It is interesting to compare the
average moments of these functions in an interval to
their counterpart in random matrices, which are the expectation values of
the characteristic polynomials of the matrix.
It turns out that these expectation values are quite interesting. For
instance, the moments of order $2K$ scale,
for  unitary invariant ensembles, as the density of eigenvalues raised to
the power $K^2$ ; the prefactor turns out to be
 a universal number, i.e. it is
independent of the specific probability distribution. An equivalent
behaviour and prefactor  had been found, as a conjecture,
within number theory. The moments of the characteristic determinants of
random matrices are computed here as limits, at coinciding points, of
multi-point correlators of determinants. These correlators are in fact
universal in Dyson's scaling limit in which the difference between the
points
goes to zero, the size of the matrix goes to infinity, and their product
remains finite.
}
\par}

\newpage

\end{titlepage}
\setlength{\baselineskip}{1.5\baselineskip}


\section{  Introduction }

   The correlation function of the eigenvalues
   of large $N\times N$ matrices are known to exhibit a number of universal
features
in the large-$N$ limit.
   For instance in the Dyson limit \cite{Mehta,Dyson}, when the distances
between these eigenvalues, measured in units of the local spacing,
 becomes of order $1/N$,  the correlation
functions, as well as the level spacing distribution,  become universal,
i.e. independent of the specific probability
   measure. For finite differences, upon a smoothing of the distribution,
the two-point
correlation function is again
   universal \cite{Ambjorn, BZ}.
   The short distance universality was also
   shown  to extend to  external source problems \cite{BH1,BH2,PZ,BHL},
   in which an  external matrix is coupled to the random matrix.

     In this article,  we study the average of the characteristic polynomials,
     whose zeros are the eigenvalues of the random matrix.
     The probability distribution of the characteristic polynomial
$\det(\la - X)$ of a random matrix $X$, a polynomial of
degree $N$ in $\la$, may be characterized by its moments
     $< \det^K( \la - X ) >$, or better by its correlation functions
$\displaystyle < \prod_{l=1}^K \det( \la_l - X ) >$.

     This study is motivated by  various conjectures which appeared
     recently in
     number theory for the zeros of the Riemann $\zeta$-function and its
generalizations known as $L$-functions \cite{KatzS}.
     Indeed the characteristic polynomials, as well as the
     zeta-fuctions,  have their zeros
     on a straight line, and these zeros obey the same statistical
     distribution.

     For the $2 K$-th moment of the Riemann $\zeta$-function ($K$ is a
     positive integer),
       it has been conjectured \cite{CG,KS} that
     \be\label{conj}
      \frac{1}{T} \int_0^T dt | \zeta( \frac{1}{2} + i t) |^{2 K}
      \simeq \gamma_K a_K  (\log T)^{K^2}
      \ee
      where $a_K$ is a number related to the Dirichlet coefficient
      (the divisor function) $d_K(n)$, and
\be \label{gamma} \gamma_K = \prod_{l=0}^{K-1}\frac{l!}{(l + K)!}.\ee
      The explicit formula  for $a_K$ is given in the Appendix, together with
      summation formulae for the Dirichlet coefficients, which are related to
      (\ref{conj}).
     In this work we shall compute the equivalent of (\ref{conj}) for
random matrices, show
that the density of states $\rho(\la)$ replaces $\log T$, and that the same
number
      $\gamma_K$  is universally present.

      For the negative moments, similar conjectures have been proposed,
      with a cut-off parameter $\delta$ for avoiding
      divergences \cite{Gonek}, and we show here how to obtain these
      negative moments for random matrices.

      Several types of  $L$-functions have been introduced \cite{KatzS},
which  correspond to the three standard classes of random matrices
      The conjecture for  the average of the  moments (\ref{conj})
      has been extended to
      these $L$-function \cite{CF}. The average is taken
      as a sum of the discriminant $d$, for instance, for the Dirichlet
      $L(\frac{1}{2},\chi_d)$ function.
      The relations between the distributions of the eigenvalues of the random
      matrix theory and the statistical distribution of the zeros
      of the various $L$-functions has also been studied
\cite{KatzS,Rubinstein}.

      Our aim in this article,  is to clarify the universality of the
      moments of the characteristic polynomials for these three classes.
       The circular unitary ensemble,
      has been studied earlier by Keating and Snaith\cite{KS}, who did
obtain the $\gamma_K$ in (\ref{gamma})
      from their calculation. However this ensemble
       has a constant density of states, and furthermore it does not allow
       to study the universality of these properties.
    In this work we have considered a Gaussian ensemble and  non-Gaussian
extensions,
      instead of the circular ensemble, to verify both the explicit
      dependence in the density of states and the universality of the
coefficient $\gamma_K$. In the process of the derivation,
	 we have found it necessary to start with the K-point functions
$\displaystyle < \prod_{l=1}^K \det( \la_l - X ) >$, which are shown to be
themselves universal in the large-N Dyson limit, in which $N(\la_i-\la_j)$
is held fixed. The moments are then simply the limit of these functions
when all the Dyson
variables vanish.

\section{  Correlation functions of characteristic polynomials}
We consider random $M \times M$ Hermitian matrices $X$ with a normalized
probability distribution
\be \label{weight} P(X) = \frac{1}{Z} \exp -N \rm{Tr} V(X), \ee
in which $V$ is a given polynomial. It will turn out to be convenient to
distinguish here $M$ and $N$, but
we later restrict ourselves to a large $N$ and large $M$ limit, with
lim$M/N=1$.
Let us consider the correlation function of $K$ distinct characteristic
polynomials :
\be F_K({\la}_1,\cdots,{\la}_K) = \langle
\prod_{{\al}=1}^{K}\det({\la}_{{\al}}-X)\rangle,\ee
in which the bracket denotes an expectation value with the weight
(\ref{weight}).

Integrating as usual over the unitary group, we obtain
\be \label{definition} F_K({\la}_1,\cdots,{\la}_K) = \frac{1}{Z_M}
\int\prod_1^M d\mu(x_i)\
\Delta^2(x_1,\cdots,x_M)
\prod_{{\al}=1}^{K}\prod_{i=1}^{M}({\la}_{\al}-x_i) \ee
in which $d\mu(x)$ denotes the measure $d\mu(x) = dx \exp-NV(x)$,  $\Delta$
the Vandermonde determinant
 $\displaystyle \Delta(x_1,\cdots,x_M) = \prod_{1\leq i<j\leq M}(x_i-x_j)$,
and $Z_M$ the normalization constant
\be \label{Z} Z_M = \int\prod_1^M d\mu(x_i)\
\Delta^2(x_1,\cdots,x_M). \ee
We now use the obvious identity
\be \Delta(x_1,\cdots,x_M)
\prod_{{\al}=1}^{K}\prod_{i=1}^{M}({\la}_{\al}-x_i) =
\frac{\Delta(x_1,\cdots,x_M;
{\la}_1,\cdots,{\la}_K)}{\Delta({\la}_1,\cdots,{\la}_K)},\ee
and represent the Vandermonde determinants $\Delta(x_1,\cdots,x_M)$ and
$\Delta(x_1,\cdots,x_M; {\la}_1,\cdots,{\la}_K)$
as determinants of arbitrary  polynomials whose coefficients of highest
degree are equal to unity (the so-called monic polynomials)
\be p_n(x) = x^n + \rm{lower degree}.\ee
Then
\be \Delta(x_1,\cdots,x_M) = \det p_n(x_m)\ee
($n$ runs from zero to $M-1$ and $m$ from one to $M$),and
\be \Delta(x_1,\cdots,x_M; {\la}_1,\cdots,{\la}_K) =\det p_a(u_b) \ee
in which $a$ runs from zero to $M+K-1$, $b$ from one to $M+K$ and $u_b$
stands for $x_b$ if $b\leq M$, or $\la_b$ for
$M<b\leq M+K$.

Choosing now the polynomials orthogonals with respect to the measure $d\mu$ :
\be \int   p_n(x)p_m(x)d\mu(x) = h_n \delta_{nm}\ ,\ee we may easily
integrate over the
$M$ eigenvalues
\be \int \prod_1^M d\mu(x_i) \ \Delta(x_1,\cdots,x_M;
{\la}_1,\cdots,{\la}_K)\Delta(x_1,\cdots,x_M) = M!
\left(\prod_0^{M-1}h_n\right)
\det p_{\al}(\la_\beta),\ee
in which $\al$ runs from $M$ to $M+K-1$ and $\beta$ from $1$ to $K$.
Similarly the normalization factor
$Z_M$ is given by
\be \label{ZM} Z_M = \int \prod_1^M d\mu(x_i) \ \Delta^2(x_1,\cdots,x_M) =
M! \left(\prod_0^{M-1}h_n\right).\ee
We thus end up with
\ba \label{F} F_K({\la}_1,\cdots,{\la}_K) =
\frac{1}{\Delta(\la_1,\cdots,\la_K)}\det \left| \begin{array}{clcr}
p_M(\la_1) &p_{M+1}(\la_1)&\cdots&p_{M+K-1}(\la_1)
\\p_M(\la_2) &p_{M+1}(\la_2)&\cdots&p_{M+K-1}(\la_2)\\ \vdots\\p_M(\la_K)
&p_{M+1}(\la_K)&\cdots&p_{M+K-1}(\la_K)\end{array} \right|.\ea
If we are concerned simply with the moments of the distribution of a single
characteristic polynomial, we obtain from (\ref{F})
\ba \mu_K(\la) =&& F_K({\la},\cdots,{\la}) = \langle \ [\det(\la-X)]^K\
\rangle\nonumber\\ =&& \frac{(-1)^{K(K-1)/2}}{\prod_{l=0}^{K-1} (l!)}\det
\left|
\begin{array}{clcr} p_M(\la) &p_{M+1}(\la)&\cdots&p_{M+K-1}(\la)
\\p'_M(\la) &p'_{M+1}(\la)&\cdots&p'_{M+K-1}(\la)\\
\vdots\\p^{(K-1)}_M(\la)
&p^{(K-1)}_{M+1}(\la)&\cdots&p^{(K-1)}_{M+K-1}(\la) \end{array}\right|.\ea

These  expressions are all exact, but in the next section we shall be
concerned with the large $N$ limit. Then (i) the interesting case is that
of even $K$,
since for odd $K$ the result is oscillatory ( for instance for $K=1$
$\mu_1(\la) = p_M(\la)$ ), (ii) it will turn out that, even if we are
interested simply in
the moments $\mu_K(\la)$, it is more convenient to study first the large
$N$-limit of the $F_K$ with distincts $\la_i$ and afterwards let
them approach a single $\la$.

The results that will be derived later for those $F_K$'s and $\mu_K$'s will
be shown to be universal in the Dyson limit, in which $N$ goes to infinity,
the
$\la_i-\la_j$ goes to zero for any pair $i,j$, and the products
$N(\la_i-\la_j)$ remain finite. We first derive explicit formulae for the
Gaussian case, and show
later that they do apply to any random matrix distribution $P(X)$ of the
form (\ref{weight}).

\section{ The Gaussian case}

We now specialize the result (\ref{F}) of the previous section to the
Gaussian distribution of $M\times M$ Hermitian matrices
\be \label{GAUSS} P(X) = \frac{1}{Z_M} \exp - \frac{N}{2} \rm Tr X^2 ,\ee
with
\be M= N-K,\ee
(the reason for this choice of $M$ will be clarified in the next section).
Then the polynomials that we have introduced, are Hermite polynomials, and
with our normalizations,
\be H_n(x) = \frac{(-1)^n}{N^n} e^{Nx^2/2} (\frac{d}{dx})^n e^{-Nx^2/2} =
x^n + \rm{l.d.},\ee
and
\be \label{h} h_n = \frac{n!}{N^n} \sqrt{\frac{2\pi}{N}}.\ee

The integral representation
\be H_n(x) = \frac{(-1)^n n!}{N^n}\oint \frac{dz}{2i\pi}
\frac{e^{-N(z^2/2+xz)}}{z^{(n+1)}}\ee
over a contour which circles around the origin in the z-plane, turns out to
be well adapted. A repeated use of this formula in the
result (\ref{F}) yields
\ba F_{2K}\label B({\la}_1,\cdots,{\la}_{2K}) &&=
\frac{(-1)^K}{\Delta(\la_1,\cdots,\la_{2K})}\frac{\prod_{l=0}^{2K-1}(M+l)!}{N^{K
(2M+2K-1)}}
\nonumber \\&& \times\oint \prod_{l=1}^{2K} \left(\frac{dz_l}{ 2i\pi
z_l^{M+l}}\right) \exp{-(N\sum_1^{2K}\frac{ z_l^2}{2})} \det (e^{-N\la_a
z_b}) .\ea
We can expand the determinant in the r.h.s. and keep only one of the
$(2K)!$ terms, antisymmetrizing instead the integration variables $z_l$.
This
gives

\ba \label{A} F_{2K}({\la}_1,&\cdots&,{\la}_{2K}) =
\frac{(-1)^K}{\Delta(\la_1,\cdots,\la_{2K})}\frac{\prod_{l=0}^{2K-1}(M+l)!}{N^{K
(2M+2K-1)}}
\nonumber \\& \times&\oint \prod_{l=1}^{2K} \left(\frac{dz_l}{ 2i\pi
z_l^{M+2K}}\right) \exp{[-N\sum_1^{2K}(\frac{ z_l^2}{2} +\la_l z_l)]} \Delta
(z_1,\cdots,z_{2K}) .\ea
This expression for the expectation value of a product of $2K$
characteristic polynomials,
as an integral over $2K$ complex variables, is exact for finite $N$ and $M$.

We are now in position to study the large $N$-limit through a saddle
point integration over each $z_l$.  Since we have chosen  $M+K=N$  each $z$ has
a weight
$\displaystyle \frac{1}{z^K}\exp-N(z^2/2 + \la z + \log z)$, which presents
two saddle points
$z_{\pm}$, solutions of the equation $z^2+\la z+1=0$, i.e. with the
parametrization
\be \la = 2 \sin \phi, \ee
when $\la$ lies on the support of the asymptotic Wigner semi-circle of the
density
of levels,
\be z_+= ie^{i\phi} , z_-= -ie^{-i\phi}.\ee
Therefore there are, a priori $2^{2K}$ saddle-points at which the moduli
of the weight $\displaystyle \exp{[-N\sum_1^{2K}(\frac{ z_l^2}{2} +\la_l
z_l+\log z_l)]}$
are the same. However, it is only when $\displaystyle \sum_1^{2K}(\frac{
z_l^2}{2} +\la_l z_l+\log z_l)$ is real (in the Dyson limit in which the
differences
between the
$\la$'s are small), that the oscillations, which damp the result, are not
present. Therefore we keep only the
$\left(\begin{array}{c}2K\\K\end{array}\right)$
saddle-points in which we take $K$ solutions of type $z_+$ and $K$ of type
$z_-$.
We are now interested in Dyson's short-distance limit. Defining
\be \la =  \frac {1}{2K}\sum_{l=1}^{2K} \la_l,\ee
and the density of eigenvalues at this point
\be \r(\la) =\frac{1}{2\pi} \sqrt{4-\la^2} = \frac{1}{\pi} \cos{\phi}, \ee
we introduced the scaling variables
\be \label{scaling} x_a = 2\pi  N \r(\la) (\la_a-\la), {\rm{with}}
\sum_{a=1}^{2K} x_a= 0, \ee
which are kept finite in this limit. Then the fluctuations around a
saddle-point may  be taken all at the point $\la$, and they yield a factor
\be \label {factor} (\frac{2\pi}{N})^K  [(1-z_+^2)(1-z_-^2)]^{-K/2}  = (N
\r(\la))^{-K} \ee
We must now take into account the various factors in (\ref{A}) at these
saddle-points. In the Dyson limit the factor $\displaystyle\prod_1^{2K}
z_l^K$ which remained in the denominator, {\it }may be
replaced by one, since at a given $\la$ one has $z_+z_-=1$. The only
delicate factor is thus
$\displaystyle\frac{\Delta(z_1,\cdots,z_{2K})}{\Delta(\la_1,\cdots,\la_{2K})}
  \exp{[-N\sum_1^{2K}(\frac{ z_l^2}{2} +\la_l z_l+\log z_l)]}  $, which we
must first compute at one of the
saddle-points, and then take the sum over the
$\left(\begin{array}{c}2K\\K\end{array}\right)$ saddle-points.
We consider first the saddle-point
\ba  z_l (\la_l) = z_+(\la_l)&   l=1,\cdots,K \nonumber\\
  z_l (\la_l) = z_-(\la_l)& l=K+1,\cdots,2K  .\ea
If we expand in the Dyson limit the weight $\displaystyle
\exp{[-N\sum_1^{2K}(\frac{ z_l^2}{2} +\la_l z_l+\log z_l)]}$ one finds
\ba  &&\exp{[-N\sum_1^{2K}(\frac{ z_l^2}{2} +\la_l z_l+\log
z_l)]}\nonumber\\
&=&\exp{NK(1+\frac{\la^2}{2})}
\times\exp{-N\left[\sum_1^K(\la_l-\la)z_+(\la)+
\sum_{K+1}^{2K}(\la_l-\la)z_-(\la)\right]},\ea
(we have used $\displaystyle \frac{d}{d\la} (\frac{1}{2}z_{\pm}^2(\la)+\la
z_{\pm} +\log z_{\pm}) = z_{\pm}$). Therefore at that saddle-point, in
terms of the
scaling variables(\ref{scaling})
\be \exp{[-N\sum_1^{2K}(\frac{ z_l^2}{2} +\la_l z_l+\log
z_l)]}=\exp{NK(1+\frac{\la^2}{2})}\exp{-i\sum_1^Kx_l}. \ee
Let us consider now the ratio of Vandermonde determinants at that same
saddle-point:
\ba \frac{\Delta(z_1,\cdots,z_{2K})}{\Delta(\la_1,\cdots,\la_{2K})}&&
=\prod_{1\leq l<m\leq K}\frac{z_+(\la_l)-z_+(\la_m)}{\la_l-\la_m}
\prod_{K+1\leq l<m\leq
{2K}}\frac{z_-(\la_l)-z_-(\la_m)}{\la_l-\la_m}\nonumber\\&&\times
\prod_{1\leq l\leq K, K+1\leq m\leq
{2K}}\frac{z_+(\la_l)-z_-(\la_m)}{\la_l-\la_m}.\ea
In the scaling limit, this factor becomes
\ba\frac{\Delta(z_1,\cdots,z_{2K})}{\Delta(\la_1,\cdots,\la_{2K})} &&=
(\frac{dz_+}{d\la}\frac{dz_-}{d\la})^{K(K-1)/2}\ (2i\cos{\phi})^{K^2}
\prod_{1\leq l\leq K, K+1\leq m\leq{2K}}\frac{1}{\la_l-\la_m}\nonumber \\
&&=(Ni)^{K^2} (2\pi\r(\la))^{K+K^2}\prod_{1\leq l\leq K, K+1\leq
m\leq{2K}}\frac{1}{x_l-x_m}.
\ea
Leaving aside for the moment the overal factors which do not change at the
various saddle-points, we note the result from this particular one
which is \\ $\displaystyle \exp{-i\sum_1^Kx_l}\prod_{1\leq l\leq K, K+1\leq
m\leq{2K}}\frac{1}{x_l-x_m}$, and consider summing  over
 the  $\left(\begin{array}{c}2K\\K\end{array}\right)$  saddle-points. The
sum is best done under the form of an integral over $K$ variables. Indeed,
if we consider
\be I(x_1,\cdots,x_{2K}) =\frac{(-1)^{K(K-1)/2}}{K!} \oint \prod_1^K \frac
{du_{\al}}{2i\pi}\exp{-i(\sum_{l=1}^K u_{\al})}\
\frac{\Delta^2(u_1,\cdots,u_K)}
{\prod_{\al=1}^K\prod_{l=1}^{2K} (u_{\al}-x_l)} \ee
over a contour in which each $u_{\al}$ circles around the $x$'s, we recover
exactly the contribution previous saddle-point by choosing $u_1=x_1,
\cdots, u_K=x_K$,
or any permutation of those $K\  x$'s. In view of the Vandermonde in the
numerator, all the $u$'s have to be different, and thus there are indeed
$\left(\begin{array}{c}2K\\K\end{array}\right)$ poles to be added, which
reconstruct exactly the sum on the saddle-points that we needed to perform..

Collecting the various factors that came on the way, we end up with the
final formula
\ba \label {final}&&
\exp-{(\frac{N}{2}\sum_{l=1}^{2K}V(\la_l))}F_{2K}(\la_1,\cdots,\la_{2K}) =
\nonumber \\&&(2\pi N\r(\la))^{K^2}
\frac{\exp(- NK)}{K!} \oint \prod_1^K \frac
{du_{\al}}{2\pi}\exp{-i(\sum_{\al=1}^K u_{\al})}\
\frac{\Delta^2(u_1,\cdots,u_K)}
{\prod_{\al=1}^K\prod_{l=1}^{2K} (u_{\al}-x_l)}. \ea
If we specialize to $K=1$ one finds
\be \label {finalx}
\exp\{-{\frac{N}{2}(V(\la_1)+V(\la_2))}\}F_{2}(\la_1,\la_{2}) = e^{-N} \
(2\pi N\r(\la)) \frac{\sin x}{x}\ee
with $x= \pi N \r(\la) (\la_1-\la_2)$, in which we recover the well-known
Dyson kernel, which characterizes the correlation between eignevalues, whose
universality has been very much discussed over the recent years. Note the
dependence in $(N\r(\la))^{K^2}$ of this function. This K=1 result
(\ref{finalx}) is indeed equal to  $(2 \pi e^{-N}) K(\la_1,\la_2)$, where
the  kernel $K(\la_1,\la_2)$
is
\be\label{Dysonkernel}
  K(\la_1,\la_2) = \frac{\sin[ \pi N \r (\la) (\la_1 - \la_2)]}{\pi (\la_1
- \la_2)}.
  \ee
(In the next section we return to the normalizations. It will be explained
how the extra-factor  $2 \pi e^{-N}$ is cancelled by the normalization
constant
$h_{N-1}$ ).

We can now specialize this formula to the moments of the distribution of
the characteristic polynomial, by letting all the $\la$'s approach each
other, i.e.
letting the $x$'s vanish. Before we do that, we should point out that the
procedure to obtain these moments is in fact subtle. In principle we could
have set
all the $\la$'s equal at an early stage of the calculation. If we returned
for instance to (\ref{B}) we might have replaced the limit of
$\displaystyle  \frac{\det (e^{-N\la_a
z_b})}{\Delta(\la_1,\cdots,\la_{2K})}$ by ${\Delta(z_1,\cdots,z_{2K})}$ (up
to a factor), but then the saddle-point
method to obtain the large $N$-limit becomes quite problematic. Indeed the
Vandermonde of the $z's$ at the saddle-point vanishes and it is necessary
to go far
beyond the Gaussian integration. However it is now straightforward to
obtain this moment from (\ref{final}).
We obtain
\ba &&\label{final2} \exp-{(NKV(\la))}F_{2K}(\la,\cdots,\la)\nonumber\\
 &=& (2\pi
N\r(\la))^{K^2}
\frac{\exp(-NK)}{K!} \oint \prod_1^K \frac
{du_{\al}}{2\pi}\exp{-i(\sum_{\al=1}^K u_{\al})}\
\frac{\Delta^2(u_1,\cdots,u_K)}
{\prod_{\al=1}^K u_{\al}^{2K}}. \ea
Expanding the Vandermonde determinant into a sum over permutations, we find
\ba\label{G39} &&\oint \prod_1^K \frac {du_{\al}}{2\pi}\exp{-i(\sum_{\al=1}^K
u_{\al})}\ \frac{\Delta^2(u_1,\cdots,u_K)}
{\prod_{\al=1}^K u_{\al}^{2K}}=
(-1)^{K(K-1)/2} \nonumber \\&&\times \sum_{P,Q} (-1)^{(P+Q)}
\frac{1}{(2K-P_0-Q_0-1)!}\cdots\frac{1}{(2K-P_{K-1}-Q_{K-1}-1)!} ,
 \ea in which $P$ and $Q$ are permutations of the integers
$(0,\cdots,K-1)$. Therefore

\ba &&\oint \prod_1^K \frac {du_{\al}}{2\pi}\exp{-i(\sum_{\al=1}^K u_{\al})}\
\frac{\Delta^2(u_1,\cdots,u_K)}
{\prod_{\al=1}^K u_{\al}^{2K}}\nonumber\\
 &=&  (-1)^{K(K-1)/2} K! \det_{0\leq i,j\leq
{K-1}}  \frac{1}{(2K-i-j-1)!}
 =K! \prod_0^{K-1}\frac{l!}{(K+l)!}, \ea
and thus finally
\be \label{moment} \exp-{(NKV(\la))}F_{2K}(\la,\cdots,\la) = (2\pi
N\r(\la))^{K^2}
e^{-NK}\prod_0^{K-1}\frac{l!}{(K+l)!} . \ee

\section{ Normalizations and Universality}
We have studied  in the previous section a Gaussian ensemble of random
matrices and found that the result (\ref{moment}) for the moment
involved $(2\pi N\r(\la))^{K^2} $ times a number and one would like to see
how general is this result, as far as the dependence in the density of
states is
concerned as well as for the normalization. We shall see that this
behaviour is quite general, and given a proper normalization, that the
prefactor is also
universal. Indeed let us recall how the K-point correlation function of the
eigenvalues are defined in an ensemble of hermitian $N\times N$ matrices
$X$ with a
probability weight proportional to $\exp{-N{\rm{Tr}}V(X)}$. In \cite{Mehta}
one finds
\ba R_K(\la_1,\cdots, \la_K) &=& \frac{N!}{(N-K)!} \frac{1}{Z_N} \int
d\la_{(K+1)}\cdots d\la_N \left\{\exp{-N\sum_1^NV(\la_l)}\right\}\nonumber\\
&\times&
\Delta^2(\la_1,\cdots\,\la_N).\ea
Comparing with our initial definitions (\ref{definition}) we see that one
has the relation
\ba\label{RK} R_K(\la_1,\cdots, \la_K)
&=& \frac{N!}{(N-K)!} \frac{Z_{N-K}}{Z_N}
\left\{\exp{-N\sum_1^KV(\la_l)}\right\}\Delta^2(\la_1,\cdots\,\la_K)
\nonumber\\
&\times&F_{2K}(\la_1,\la_1,\cdots,\la_K,\la_K) ;
\ea
the r.h.s. reduces, up to a normalization,  to our previous product of
characteristic functions of matrices $(N-K)\times (N-K)$, each one beeing
repeated twice. On the other hand
it is well
known (\cite{Mehta}) that this K-point function may be expressed in terms
of a kernel $K_N$ as
\be\label{DET} R_K(\la_1,\cdots, \la_K) = \det_{1\leq i,j\leq K}
K_N(\la_i,\la_j) ,\ee
and without entering into the precise definition
of $K_N$ in terms of orthogonal polynomials, one
should simply recall that
$\displaystyle \frac{K_N(\la,\mu)}{\r((\la + \mu)/2))}$
is  universal in the Dyson limit(\cite{BZ}) ($\la-\mu$ goes to zero, N goes
to infinity, $N(\la-\mu)$ finite), i.e. it is independent of the polynomial
$V$ which
defines the probability measure.

Therefore we define a modified weight, and modified moments,

\be \Phi_{2K}(\la_1,\la_2,\cdots,\la_{2K})=\frac{N!}{(N-K)!}\frac{Z_{N-K}}{Z_N}
\left\{\exp{-\frac{N}{2}\sum_1^{2K}V(\la_l)}\right\}
F_{2K}(\la_1,\la_2,\cdots,\la_{2K}) \ee and
\be M_{2K}(\la) = \frac{N!}{(N-K)!}\frac{Z_{N-K}}{Z_N}
\left\{\exp{-NKV(\la)}\right\}F_{2K}(\la,\la,\cdots,\la).\ee
The universality of level correlations implies the universality of
$M_{2K}$. Therefore we have to return to the Gaussian case, in order to
take into account this
new normalization, and then the result will be universal.

>From (\ref{ZM}) we have
\be  \frac{N!}{(N-K)!}\frac{Z_{N-K}}{Z_N} =\frac{1}{\prod_{N-K}^{N-1}h_n}, \ee
and, given the explicit expression (\ref{h}) of $h_n$ for the Gaussian
case, we find, in the large N limit,
\be  \frac{N!}{(N-K)!}\frac{Z_{N-K}}{Z_N} = (2\pi)^{-K} e^{NK}. \ee
With this normalization the universal moment $M_{2K}(\la)$ is given by
\be \label{M} M_{2K}(\la) = (2\pi)^{-K} (2\pi N\r(\la))^{K^2}
\prod_0^{K-1}\frac{l!}{(K+l)!}\ee

In fact this connection between the usual correlation functions
and the expectation values of a product of characteristic functions,
(\ref{RK}) and (\ref{DET}),
allows one to recover directly  the moment
$M_{2K}(\la)$, by using the universal expression
for the  kernel
$K(\la_i,\la_j)$ in the Dyson
limit,
\be
K(\la_i,\la_j) = \frac{\sin [\pi N \r (\la_i - \la_j)]} {\pi (\la_i - \la_j)}.
\ee
The integral representation, over $2K$ variables describing contours around
the $K$ poles
$\lambda_l$,
\be
\frac{\det_{1\leq i<j\leq K} K(\la_i,\la_j)}{\Delta^2(\la_1,\cdots,\la_K)} =
\frac{1}{K!} \oint \prod_1^K\frac{du_l}{2\pi i}\oint
\prod_1^K\frac{dv_l}{2\pi i}
\frac{\Delta (u_1,\cdots,u_K)\Delta
(v_1,\cdots,v_K)}{\prod_{i=1}^K\prod_{j=1}^K (u_i - \la_j)
 (v_{i} - \la_{j})} \prod_{i=1}^K
K(u_i,v_i)
\ee
allows one to write easily the limit in which all the $\la$'s are equal:
\ba
{\rm { lim}}\frac{\det_{1\leq i<j\leq K}
K(\la_i,\la_j)}{\Delta^2(\la_1,\cdots,\la_K)} &=&
\frac{1}{K!} \oint \prod_1^K\frac{du_l}{2\pi i}\oint
\prod_1^K\frac{dv_l}{2\pi i}
\frac{\Delta (u_1,\cdots,u_K)\Delta (v_1,\cdots,v_K)}{\prod_{i=1}^K (u_i -
\la)^K
(v_{i} - \la)^K }\nonumber\\
&\times&\prod_{i=1}^K
K(u_i,v_i)
\ea
Since the kernel is a Toeplitz matrix, i.e. $K(\la_i,\la_j) = K(\la_i -
\la_j)$, one can shift the
$u$'s and the $v$'s of $\la$ and
the r.h.s.  becomes independent of $\la$.
In the case of the sine kernel we obtain, in the limit in which all the
$\lambda$'s are equal,
\ba
&&\frac{1}{K!} \oint \prod_1^K\frac{dx_l}{2\pi i} \oint
\prod_1^K\frac{dv_l}{2 \pi i}
\frac{\Delta(v_1,\cdots,v_K)
\Delta(x_1,\cdots,x_K)}{\prod_{i=1}^K [(v_i + x_i)^K v_i^K]}\prod_{i=1}^K
\frac{\sin(\pi N
\r x_i)}{\pi x_i}\nonumber\\
&=& \frac{(2 \pi \r N)^{K^2}}{(2 \pi)^K} \prod_{l=0}^{K-1}\frac{l!}{(l + K)!}.
\ea
We have indeed recovered, for any function $V$ defining the probability
distribution,
the universal moment (\ref{M})

\section{ Large N asymptotics}

Rather than starting, as in the previous sections, of exact expression for
the correlation functions of characteristic functions, and at the end
letting $N$ go to infinity,
we may use a different method to investigate directly
the large N limit for the moments of their distribution. This method
applies for
a  general probability distribution of the form (\ref{weight}) and it may
also be used to the more general
case of an external matrix source
 coupled to the matrix $X$ \cite{PZ} in this distribution.
It turns out that here again it is neccessary to consider first $F_{2K}$
for different $\la_j$'s,
and let go all the $\la_j$'s approach the same  $\la$ at the end of the
calculation.

>From (\ref{definition}), we have
\be
\frac{\partial \ln F_{2K}}{\partial \la_i} = M G_{\la}(\la_i)
\ee
where $G_\la(\la_i)$ is the resolvent,
\be G_\la(\la_i) = \frac{1}{M} < \Tr \frac{1}{\la_i - X} >.
\ee
The bracket here denotes an expectation value with a weight which includes
both $P(X)$ and
$\displaystyle \prod_1^{2K} \det(\la_l-X)$.
We assume that the asymptotic spectrum of the eigenvalues $x_i$ of $X$ fill
a single interval $[\alpha,\beta]$
in the large M limit. (It is sufficient to consider the single cut case,
since we are
interested in  Dyson short distance universality, which involves only
the local statistics).
Therefore $G_\la(z)$ is also analytic in a  plane cut from the interval
$[\alpha,\beta]$, and
\be
G_\la(x\pm i 0) = \hat G_\la(x) \mp i \pi \r_\la(x) \ee
where $\hat G_\la(x) = [G_\la(x+i 0) + G_\la(x - i 0)]/2$.
The saddle point equation in the large M limit becomes
\be \label{SP}2 M G_{\la}(z) - N V'(z) + \sum_{j=1}^{2K} \frac{1}{z - \la_j
} = 0
.\ee
The last term of (\ref{SP}) is of relative order $1/N$ and thus
we have to solve this Riemann-Hilbert problem to this order. At leading
order, we have $2 \hat G(x) = V'(x)$, and up to
order $1/N$,
\be
G_\la(z) = G(z) + \frac{1}{N}(C_G(z) + \sum_{i=1}^{2K} C_{\la_i}(z)).
\ee

>From the saddle point equation (\ref{SP}), we have $\hat C_G(x) = (N - M)
\hat G(x)$
and $\displaystyle \hat C_{\la_i}(x) = \frac{1}{2(\la_i - x)}$.
We now set $M= N - K$. The  functions  $C_G(z)$ and
$C_{\la_i}(z)$ are uniquely determined from their analyticity in a plane
cut from $\alpha$ to $\beta$, and their fall-off as $1/z^2$ for  large $z$
(since both
$G_{\la}(z)$ and $G(z)$ behave as $1/z$ at infinity). The result is
\ba
C_G(z) &=& K G(z) - {K\over{\sqrt{(z - \alpha)(z - \beta)}}}\nonumber\\
C_{\lambda_i}(z) &=& {1\over{2}} {1\over{\sqrt{(z - \alpha)(z - \beta)}}}
( 1- {\sqrt{(z - \alpha)(z - \beta)} - \sqrt{(\lambda_i - \alpha)(\lambda_i -
\beta)}\over{
z - \lambda_i}} )\nonumber\\
\ea
These expressions lead to
\ba
(N - K) G_{\lambda}(\lambda_i) &=& N G(\lambda_i)
- {1\over{2}}{d\over{d \lambda_i}}
\log \sqrt{(\lambda_i - \alpha)(\lambda_i - \beta)}\nonumber\\
&-& {1\over{2}} \sum_{j=1,j\ne i}^{2 K}{1\over{\lambda_i - \lambda_j}}
( 1 - \sqrt{{(\lambda_j - \alpha)(\lambda_j - \beta)\over{(\lambda_i - \alpha)(
\lambda_i - \beta)
}}} ).
\ea
 Since there
is a
branch cut between $\alpha$ and $\beta$,  one must specify
 whether
$\lambda_i$ approaches the real axis from above or from below.
The sign of the square root on both sides of the cut will be denoted
$\epsilon_{i}$.
There are then a priori $2^{2 K}$ saddle points corresponding to the
different choices of
$\epsilon_{i}$.
For each choice of the $\epsilon_i$'s, we have
\ba
{\partial\over{\partial \lambda_i}} \log \tilde F_{\epsilon} &=&
\epsilon_{i} N i \pi \rho(\lambda_i) +
- {1\over{2}}{d\over{d \lambda_i}}
\log \sqrt{(\lambda_i - \alpha)( \beta - \lambda_i)}\nonumber\\
&-& {1\over{2}} \sum_{j=1,j\neq i}^{2 K}{1\over{\lambda_i - \lambda_j}}
( 1 - {\epsilon_{j}\sqrt{(\lambda_j - \alpha)( \beta - \lambda_j)}
\over{\epsilon_i\sqrt{(\lambda_i - \alpha)(\beta - \lambda_i)}}}
 )
\ea
where $\tilde F_{\epsilon}$ means the value of $F_{2K}$ for given
$\epsilon_j$'s multiplied a factor $\exp( - \frac{N}{2}\sum V(\la_i))$.
Introducing the parametrization $\phi(x)$,defined by $
x = {1\over{2}}(\alpha + \beta) - {1\over{2}} (\beta - \alpha) \cos \phi(x)$
and ${1\over{2}}(\beta - \alpha) \sin \phi(x) =
\sqrt{(x - \alpha)(\beta - x)}$, we have
\ba
&&{d\over{d\lambda_i}} \log \sin ({\epsilon_i \phi(\lambda_i) - \epsilon_j
\phi(\lambda_j)
\over{2}})\nonumber\\
 &=& {\epsilon_i\over{2}}{1\over{\sqrt{(\lambda_i - \alpha)(\beta - \lambda_i)}
}}{\epsilon_i \sqrt{(\lambda_i - \alpha)(\beta - \lambda_i)} + \epsilon_j
\sqrt{(\lambda_j - \alpha)(\beta - \lambda_j)}\over{\lambda_i - \lambda_j}}
\nonumber\\
\ea
Thus we obtain $\tilde F_{\epsilon}$ by integration,
\ba
\tilde F_{\epsilon} &=& C_{\epsilon} \prod_{i<j}^{2 K}
{\sin({\epsilon_i\phi(\lambda
_i)
-\epsilon_j\phi(\lambda_j)\over{2}})\over{\lambda_i - \lambda_j}}
\prod_{i=1}^{2 K}{1\over{\sqrt{\sin \phi(\lambda_i)}}}
\nonumber\\
&\times& \prod_{i=1}^{2 K}
\exp( \epsilon_i i N \pi \int_{\lambda_0}^{\lambda_i} \rho(x) dx )
\ea
We have to sum over all the saddle-point contributions, i.e. sum over all
the different
choices of $\epsilon_j$'s.
We focus now on the Dyson limit in which  the differences $\lambda_i -
\lambda_j$ are all of order 1/N.
Among the $2^{2K}$ possibilities, we retain only the
$\left(\begin{array}{c}2K\\K\end{array}\right)$ solutions
in which  half of
$K$ among $\epsilon_l$ are positive, and the remaining halves are negative.
Otherwise, the
exponential factor in the final result gives very rapid oscillations in the
large N limit. This situation is
thus exactly similar to that of the previous section.

Again the sum over the $\left(\begin{array}{c}2K\\K\end{array}\right)$
saddle-points is conveniently
written as a contour
integral
 \be   \tilde F = {1\over{K!}}\oint \cdots \oint {du_1 du_2 \cdots
du_K\over{(2 \pi i)^K}}
   {\prod_{n<m} (u_n - u_m)^2\over{\prod_{n=1}^{K}\prod_{j=1}^{2 K}
   (u_n - \lambda_j)}}\cos(\sum_{j=1}^{2 K}\lambda_j - 2\sum_{n=1}^K u_n)
\ee
When we set all the $\lambda_j= \lambda$, this becomes
\be
   \tilde F = {1\over{K!}}\oint {\prod du_i\over{(2\pi i)^{K}}}{\prod_{i<j}
(u_i - u_j)^2
  \over{\prod_{i=1}^K u_i^{2 K}}}
  \cos(2\sum_{n=1}^K u_n \pi N \r)
 \ee
and we recover the result (\ref{final2}).
However in this method, since we re-integrated the logarithmic derivative
of $F_{2K}$, the constant of integration
remains  undetermined. We may fix this constant by the same requirement
that we have
used in the previous section, and the final result agrees then with
the previous calculation.

\section{Symplectic group $Sp(N)$ }
 We have studied up to now unitary invariant measures, characterized for
the probability law of the eigenvalues by the
factor $\vert\D(x_1,\cdots,x_M)\vert^\beta$. We could also consider the
Gaussian orthogonal ensemble (GOE, with $\beta=1$) or
Gaussian symplectic (GSE, with $\beta=4$). If we took the GOE for instance,
we could immediately relate the
correlation functions of characteristic determinants, to the correlations
of the eigenvalues, as in (\ref{RK}) (except that since $\beta$ is one
no doubling of the $\la$'s is needed), and therefore relate the moments
universality to the Dyson universal limit.
Remaining still with the unitary $\beta=2$ class, in Cartan's
classification of symmetric spaces, we find ensembles which are invariant
under
$Sp(N)$ or $O(N)$.
   One of the physical applications of random  $Sp(N)$ matrices, is the
statistics of
   the energy levels inside a superconductor vortex \cite{BHL}.
   In  number theory, it is known that some generalizations of Riemann's
$\zeta$-functions,
such as Dirichlet $L$-function $L(s,\chi_d)$ where $\chi_d$ is
a quadratic  Dirichlet character of  mod $|d|$,
   present a spectrum of  low lying zeros on the line ${\rm{Re}}\  s=1/2$,
which  agrees with the statistics of the eigenvalues
 of the Sp(N) random matrix theory
   \cite{KatzS,Rubinstein}. In this Sp(N) invariant symmetric spaces,
   the eigenvalues appear alway in pairs of positive and negative real numbers.
   Due to this fact, a new universality class governs the correlations of
the eigenvalues near the origin, i.e. near $s=1/2$, (whereas in the bulk one
recovers the  previous unitary class).

   Therefore we study now the new universality class, which governs  the
new scaling
near the origin.
   We thus consider random Hermitian matrices $X$, which are $2M\times 2M$
and  satisfy the condition
\be
 X^{T} J + J X = 0
\ee
where $J$ is
\be
 J = \left(\matrix{ 0 & 1_M\cr
-1_M & 0} \right).
\ee
The unitary symplectic group is a the subgroup of $SU(2M)$ consisting of
$2M\times 2M$ unitary matrices, satisfying the symplectic constraint
\be U^T = -JU^{\dagger} J\ee
   The integration over this unitary symplectic group for
$F_K(\la_1,\cdots,\la_K)$
   gives \cite{BHL}
   \ba\label{definition2} F_K({\la}_1,\cdots,{\la}_K)
   &=& < \prod_{\alpha=1}^K \det (\la_{\al} - X) >\nonumber\\
   &=& \frac{1}{Z_M}
\int\prod_1^M d\mu(x_i)\
\Delta^2(x_1^2,\cdots,x_M^2)\prod_{i=1}^M x_i^2
\prod_{{\al}=1}^{K}\prod_{i=1}^{M}({\la}_{\al}^2-x_i^2) \ea
   Repeating the analysis of section 2, $F_K(\la_1,\cdots,\la_K)$ is
   given again by a determinantal form as (\ref{F}).
    Changing $x_1$ to $x_i^2 = y_i$ and denoting $\mu_i = \la_i^2$, we have
    \be
    F_K(\mu_1,\cdots,\mu_K) =  \int_0^{\infty} \prod_{i=1}^K dy_i \prod_{i=1}^K
     y_i^{\frac{1}{2}} \prod_{i<j}(y_i - y_j)^2 \prod_{\al=1}^K \prod_{i=1}^K
     (\mu_\al - y_i) e^{-N\sum y_i}
    \ee
    The orthogonal monic polynomials for this measure are the Laguerre
polynomials
    $L_n^{(\frac{1}{2})}(y)$, which is  defined by
    \ba
    L_n^{(\frac{1}{2})}(y) &=& \frac{(-1)^n}{\sqrt{y}} \frac{e^{Ny}}{N^n}
    (\frac{d}{dy})^n (y^{n + \frac{1}{2}} e^{- N y})\nonumber\\
    &=& \frac{(-1)^n}{N^n} n! \oint \frac{du}{2 \pi i}
    \frac{(1 + u)^{n + \frac{1}{2}}}{u^{n+1}} e^{- N u y}
    \ea
    normalized as required to  $L_n^{(\frac{1}{2})}(y) = y^n + {\rm{lower
degree}}$. The orthogonality condition is
    \be
    \int_0^{\infty} dy e^{- N y}\sqrt{y} L_n^{(\frac{1}{2})}(y)
L_m^{(\frac{1}{2})}(y)
    = h_n\delta_{n,m}
    \ee
    with $h_n = n! \Gamma(n + \frac{3}{2})/N^{2n + \frac{3}{2}}$, and $h_{N-1}
    \simeq 2\pi  e^{- 2N}$ in the large N limit.

     From (\ref{F}), we have
     \ba
     &&F_K(\mu_1,\cdots,\mu_K) = (-1)^{K(M + \frac{K-1}{2})}
     \frac{\prod_{l=0}^{K-1}
     (M + l)!}{N^{K(M + \frac{K}{2} -
\frac{1}{2})}}\frac{1}{\Delta(\mu)}\nonumber\\
     &\times& \oint \prod_{i=1}^K (\frac{dz_i}{2 \pi i} )
     \prod_{l=1}^K \frac{(1 + z_l)^{M + K -
     \frac{1}{2}}}{z_l^{M+K}}e^{- N \sum z_\al \mu_\al} \prod_{i<j}^K
     ( \frac{z_i}{1 + z_i} - \frac{z_j}{1 + z_j})
     \ea
     We now set $M = N - K$, and the factor $\displaystyle \prod_0^{K-1} (M
+ l)!/N^{K(M+K/2 -1/2)}$ is equal to
     $(2 \pi N)^K e^{- K N}$ , up to corrections of relative order $1/N$ in
the large N limit.
     The  large N limit is governed by the saddle-point equations
      $z_l^2 + z_l + \frac{1}{\mu_l} = 0$. In the following we study the
scaling vicinity of the origin, in which all the $\mu_l$'s
     scale as $1/N^2$. Then $z_l^2$ at the saddle-point may be expanded
     \be\label{saddle}
     z_l \simeq  \frac{i \epsilon_l}{\sqrt{\mu_l}} - \frac{1}{2} +
O(\sqrt{\mu_l})
     \ee
     where $\epsilon_l = \pm 1$.

     Noting that $\displaystyle \prod_{i<j} (\frac{z_i}{1 + z_i} -
\frac{z_j}{1 + z_j}) =
     \prod_{i<j} ( - z_i^2 \mu_i + z_j^2 \mu_j) \simeq i^{K(K-1)/2}\prod
(\epsilon_i
     \la_i - \epsilon_j
     \la_j)$ ($\epsilon_i=\pm1$),
     and combining it with the Vandermonde $\Delta(\la^2)$, we are left with
     a factor $\prod_{i<j}\frac{1}{\epsilon_i \la_i + \epsilon_j \la_j}$
      in this scaling limit.
       We have also the exponential $e^{- N \sum z_\al \mu_\al} = e^{ - i \sum
       \epsilon_i \la_i}$.

We have again to sum over all the saddle-points, which are
       characterized by the sign of $\epsilon_i= \pm 1$, and to include the
factor due to the
fluctuations near the saddle-point. The Gaussian fluctuations yield a factor
       $(2\pi/(-2 i \epsilon_i \la_i^3 N))^{1/2}$.  Then
        $(1 + z_l)^{-1/2} \simeq (\la_i/(\epsilon_i i))^{1/2}$.
      There is a $1/(2\pi i)^K$  in addition. We have an extra
      $\epsilon_i$ due to the contour direction, which goes through two saddle
      points ; one is in the positive imaginary plane and the other in the
negative
      half-plane.
     When K=2, and $\la_1$ and $\la_2$ are of order $1/N$, we obtain

     \be
          F_2(\la_1,\la_2) = \frac{2 \pi e^{-2N}}{\la_1 \la_2}
K_{SP}(\la_1,\la_2),
     \ee

    with the kernel
      $K_{SP}(\la_1,\la_2)$ given by,
     \be\label{kernel2}
     K_{SP}(\la_1,\la_2) = \frac{\sin[N(\la_1 - \la_2)]}{2 \pi (\la_1 - \la_2)}
     - \frac{\sin[N(\la_1 + \la_2)]}{2\pi(\la_1 + \la_2)}
     \ee
     The coefficient $ (2 \pi)e^{-2N}$ is cancelled by the normalization
     factor $1/h_{N-1}$.
     Putting $\la_1=\la_2 = 0$, we have
     neglecting the factor  $2 \pi e^{-2N}$, $F_2(0) \simeq \frac{1}{2\pi}
     \frac{4}{3!} N^3$.

    For general K, $F_K(\la_1,\cdots,\la_k)$ becomes in the scaling  limit
     \ba
       F_K(\la_1,\cdots,\la_K) &=& (-1)^{K(N-K + \frac{K-1}{2})}
       (2 \pi N)^{\frac{K}{2}}
       e^{-NK}(i)^{\frac{K}{2} (K-1)} (\frac{\pi}{N})^{\frac{K}{2}}
       \frac{1}{(2 \pi i)^K}
       \nonumber\\
       &\times&\sum_{\epsilon}
         \frac{e^{-i N \sum_i \epsilon_i
       \la_i}}{\prod_{i=1}^K \epsilon_i \la_i
       \prod_{i<j} (\epsilon_i \la_i + \epsilon_j \la_j)}
       \ea
     The sum over all the saddle-points, characterized by $\epsilon_i \pm 1$,
      is conveniently written as a  contour integral,
     \ba\label{spsum}
     I &=& \sum_{\epsilon}
     \frac{1}{\prod_{i<j}(\epsilon_i \la_i + \epsilon_j \la_j)\prod (\epsilon_i
     \la_i)}
     e^{-i N \sum \epsilon_i \la_i}\nonumber\\
     &=& (-1)^{\frac{K}{2} (K - 1)}{2^k\over{k!}}\oint \cdots \oint
\prod_{i=1}^k
     ({du_i\over{2\pi i}}) {\Delta(u^2) \Delta(u)\over{\prod_{i=1}^k
     \prod_{j=1}^k (u_i^2 - \lambda_j^2)}} e^{-i N (\sum_{i=1}^k u_i )}
     \ea
     where the contour encloses $u_i = \pm \la_j$. We may now set  $\la_j =
\la$,
     and keeping track of  various coefficients, we obtain the K-th moment
     $F_K(\la,\cdots,\la)$. For  general $\la$, the result has a
complicated form, but when
     $\la=0$, it becomes a number

     \ba\label{spfinal1}
     \ F_K(0,\cdots,0) &=& {2^{k/2}e^{-NK}\over{k!}}N^{\frac{K}{2}(K + 1)}
     (i)^{\frac{K}{2} (K - 3)}(-1)^{K(N-1)}\nonumber\\
     &\times& \oint  \prod_{i=1}^k
     ({du_i\over{2\pi i}}) \frac{\Delta(u^2) \Delta(u)}{\prod_{i=1}^k
      u_i^{2 K}} e^{- i  \sum_{i=1}^k u_i }
     \ea
     This representation allows one to compute the K-th moment at the origin.
     By the Expansion of the VanderMonde determinants, similarly to
(\ref{G39}),
     (\ref{spfinal1}) is reduced to a determinant form. We have by the
normalization; $\tilde F_K(0) = (2 \pi )^{- \frac{K}{2}}
     e^{KN} F_K(0)$,
     \be\label{spfinal2}
       \tilde F_K(0) = (- 1)^{K N} \prod_{l=1}^K \frac{l!}{(2 l)!}
       \frac{(2N)^{\frac{K}{2}(K + 1)}}{\pi^{\frac{K}{2}}}
     \ee
       Comparing to the result of the unitary case in (\ref{M}),
       we notice that the exponent of N is different and the universal
       coefficient is given also
       by the product of the ratio of the factorizations.

      For $F_{2K}(\la_1,\la_1,\cdots,\la_K,\la_K)$, the even 2K-th moment
may be obtained again from
     $\tilde F_{2K}(\la_1,\la_1,\cdots,\la_K,\la_K)
      =   \det[K(\la_i,\la_j)]/(\Delta^2(\la^2)\prod \la_i^2)$ ;
     using the expression for the kernel (\ref{kernel2}), we have
     for the 2K-th moment,
     \ba
     \frac{\det[K(\la_i,\la_j)]}{\Delta^2(\la^2)\prod \la_i^2} &=&
     \frac{2^K}{K!} \oint \prod \frac{du_i}{2 \pi i}
     \oint \prod \frac{dv_i}{2\pi i}
     \frac{\Delta(u^2)\Delta(v^2)}{\prod_{i=1}^K\prod_{j=1}^K(u_i^2 - \la_j^2)
     \prod_{i=1}^K\prod_{j=1}^K (v_i^2 - \la_j^2)}\nonumber\\
     &\times& \frac{1}{(2\pi)^K}
     \prod_{i=1}^K \frac{\sin[  N(u_i - v_i)]}{u_i - v_i}.
     \ea
     For general $\la$, the result has a complicated form, but again here
one can compute form there
the values at $\la=0$. The result agrees with the previous expression of
$\tilde F_{2K}(0)$ in (\ref{spfinal2}).
     One may also use the large N  asymptotic analysis
      as in  section 5 and rederive the  results as the same sum (\ref{spsum})
      over the saddle points.

 \section{Orthogonal group O(N)}

    We discuss here the $O(2N)$ case, which is different from $Sp(N)$ (
whereas $O(2N + 1)
    $ has a structure which is similar to $Sp(N)$ \cite{BHL}).
In number theory, for example the twisted $L$-function,
$L_{\tau}(s,\chi_d)$ presents a spectrum of low lying zeros,
which agrees with the statistics
of the eigenvalues of the $O(2N)$ random matrix theory \cite{KatzS,Rubinstein}.
Then in terms of eigenvalues
       \ba\label{definition3} F_K({\la}_1,\cdots,{\la}_K)
   &=& < \prod_{\alpha=1}^K \det (\la_{\al} - X) >\nonumber\\
   &=& \frac{1}{Z_M}
\int\prod_1^M d\mu(x_i)\
\Delta^2(x_1^2,\cdots,x_M^2)
\prod_{{\al}=1}^{K}\prod_{i=1}^{M}({\la}_{\al}^2-x_i^2) \ea
   The difference between the symplectic and orthogonal case is due to the
absence of
   the factor $\prod x_i^2$.
   Using the  analysis of section 2, $F_K(\la_1,\cdots,\la_K)$ is
   given by the determinantal form as in (\ref{F}).
   Changing $x_i^2 $ to $ y_i$ and denoting $\mu_i = \la_i^2$, we have
    \be
    F_K(\mu_1,\cdots,\mu_K) =  \int_0^{\infty} \prod_{i=1}^K dy_i \prod_{i=1}^K
     y_i^{-\frac{1}{2}} \prod_{i<j}(y_i - y_j)^2 \prod_{\al=1}^K \prod_{i=1}^K
     (\mu_\al - y_i) e^{-N\sum y_i}
    \ee
    The orthogonal polynomials for this case are Laguerre polynomials
    $L_n^{(-\frac{1}{2})}(y)$, which is  defined by
    \ba
    L_n^{(-\frac{1}{2})}(y) &=& (-1)^n \sqrt{y} \frac{e^{Ny}}{N^n}
    (\frac{d}{dy})^n (y^{n - \frac{1}{2}} e^{- N y})\nonumber\\
    &=& \frac{(-1)^n}{N^n} n! \oint \frac{du}{2 \pi i}
    \frac{(1 + u)^{n - \frac{1}{2}}}{u^{n+1}} e^{- N u y}
    \ea
   normalized as  $L_n^{(- \frac{1}{2})}(y) = y^n + {\rm{lower degree}}$.
    The orthogonality condition is
    \be
    \int_0^{\infty} dy e^{- N y}\frac{1}{\sqrt{y}} L_n^{(-\frac{1}{2})}(y)
L_m^{(-\frac{1}{2})}(y)
    = h_n\delta_{n,m}
    \ee
    with $h_n = n! \Gamma(n + \frac{1}{2})/N^{2n + \frac{1}{2}}$, and $h_{N-1}
    \simeq 2\pi  e^{- 2N}$ in the large N limit.
     From (\ref{F}), we have similar to the $Sp(N)$ case,
     \ba
     F_K(\mu_1,\cdots,\mu_K) &=& (-1)^{K(M + \frac{K-1}{2})}
     \frac{\prod_{l=0}^{K-1}
     (M + l)!}{N^{K(M + \frac{K}{2} -
\frac{1}{2})}}\frac{1}{\Delta(\mu)}\nonumber\\
     &\times& \oint \prod_{i=1}^K (\frac{dz_i}{2 \pi i} )
     \prod_{l=1}^K \frac{(1 + z_l)^{M + K -
     \frac{3}{2}}}{z_l^{M+K}}e^{- N \sum z_\al \mu_\al} \prod_{i<j}^K
     ( \frac{z_i}{1 + z_i} - \frac{z_j}{1 + z_j})\nonumber\\
     \ea
     We  set $M = N - K$, and the factor $\displaystyle \prod_0^{K-1} (M
+ l)!/N^{K(M+K/2 -1/2)}$ is equal to
     $(2 \pi N)^K e^{- K N}$ , up to corrections of relative order $1/N$ in
the large N limit.
      The saddle point $z_l$ is same as (\ref{saddle}). The only difference
is the
      extra factor $(1 + z_l)^{-1} \simeq \frac{\la_l}{i \epsilon_l}$.
       When K=2,  we obtain
     \be
          F_2(\la_1,\la_2) = 2 \pi e^{-2N} K_{O}(\la_1,\la_2),
     \ee
     with the kernel
      $K_{O}(\la_1,\la_2)$ given by,
     \be\label{kernel3}
     K_{O}(\la_1,\la_2) = \frac{\sin[N(\la_1 - \la_2)]}{2 \pi (\la_1 - \la_2)}
     + \frac{\sin[N(\la_1 + \la_2)]}{2\pi(\la_1 + \la_2)}
     \ee
     The factor $ (2 \pi)e^{-2N}$ is cancelled by the normalization
     factor $1/h_{N-1}\simeq (2\pi )^{-1}e^{2N}$.
     Putting $\la_1=\la_2 = 0$, we have
     neglecting the factor  $2 \pi e^{-2N}$, $F_2(0) \simeq \frac{1}{2\pi}
     (2 N)$.

    For general K, $F_K(\la_1,\cdots,\la_k)$ becomes in the scaling  limit
     \ba
       F_K(\la_1,\cdots,\la_K) &=& (-1)^{K(N-K + \frac{K-1}{2})}
       (2 \pi N)^{\frac{K}{2}}
       e^{-NK}(i)^{\frac{K}{2} (K-3)} (\frac{\pi}{N})^{\frac{K}{2}}
       \frac{1}{(2 \pi i)^K}
       \nonumber\\
       &\times&\sum_{\epsilon}
         \frac{e^{-i N \sum_i \epsilon_i
       \la_i}}{
       \prod_{i<j} (\epsilon_i \la_i + \epsilon_j \la_j)}
       \ea
     The sum over all the saddle-points, characterized by $\epsilon_i \pm 1$,
      is conveniently written as a  contour integral,
     \ba\label{sum3}
     I &=& \sum_{\epsilon}
     \frac{1}{\prod_{i<j}(\epsilon_i \la_i + \epsilon_j \la_j)\prod (\epsilon_i
     \la_i)}
     e^{-i N \sum \epsilon_i \la_i}\nonumber\\
     &=& (-1)^{\frac{K}{2} (K - 1)}{2^k\over{k!}}\oint \cdots \oint
\prod_{i=1}^k
     ({du_i\over{2\pi i}}) {\Delta(u^2) \Delta(u) \prod_{i=1}^K u_i
     \over{\prod_{i=1}^k
     \prod_{j=1}^k (u_i^2 - \lambda_j^2)}} e^{-i N (\sum_{i=1}^k u_i )}
     \ea
     where the contour encloses $u_i = \pm \la_j$. We may now set  $\la_j =
\la$,
     and keeping track of  various coefficients, we obtain the K-th moment
     $F_K(\la,\cdots,\la)$. For  general $\la$, the result has a
complicated form, but when
     $\la=0$, it becomes a number
\ba\label{onfinal}
     \ F_K(0,\cdots,0) &=& {2^{k/2}e^{-NK}\over{k!}}N^{\frac{K}{2}(K - 1)}
     (i)^{\frac{K}{2} (K - 5)}(-1)^{K(N-1)}\nonumber\\
     &\times& \oint  \prod_{i=1}^k
     ({du_i\over{2\pi i}}) \frac{\Delta(u^2) \Delta(u)}{\prod_{i=1}^k
      u_i^{2 K - 1}} e^{- i  \sum_{i=1}^k u_i }
     \ea
     The normalization factor is $(2\pi)^{-\frac{K}{2}} e^{KN}$ for
     $F_K(\la)$. Denoting the normalized K-th moment by $\tilde F_K(\la)$,
     we have
     \be\label{onfinal2}
       \tilde F_K(0) = (-1)^{KN} \prod_{l=1}^{K-1} \frac{l!}{(2 l)!}
       \frac{(2N)^{\frac{K}{2}(K-1)}}{\pi^{\frac{K}{2}}}
     \ee
     We have
      $\tilde F_{2K}(\la_1,\la_1,\cdots,\la_K,\la_K)
      =   \det[K(\la_i,\la_j)]/\Delta^2(\la^2)$.
     Using the expression for the kernel (\ref{kernel3}), we obtain
     for the 2K-th moment in the orthogonal $O(2N)$ case,
     \ba
     \frac{\det[K(\la_i,\la_j)]}{\Delta^2(\la^2)} &=&
     \frac{2^K}{K!} \oint \prod \frac{du_i}{2 \pi i}
     \oint \prod \frac{dv_i}{2\pi i}
     \frac{\Delta(u^2)\Delta(v^2) \prod_{i=1}^K (u_i
v_i)}{\prod_{i=1}^K\prod_{j=1}^K(u_i^2 - \la_j^2)
     \prod_{i=1}^K\prod_{j=1}^K (v_i^2 - \la_j^2)}\nonumber\\
     &\times& \frac{1}{(2\pi)^K}
     \prod_{i=1}^K \frac{\sin[  N(u_i - v_i)]}{u_i - v_i}.
     \ea
     Inserting $\la_i =0$, we find the consistent result with (\ref{onfinal}).

\section {Negative moments}

 In the number theory literature one finds various moments in which powers
of the zeta-functions appear in the denominator  \cite{Gonek}.
The equivalent for random matrices would be to consider expectations values
of the form $<\displaystyle \prod_1^K\det(\la_l-X)^{\epsilon_l}>$
in which the $\epsilon$'s are $\pm 1$. One cannot use any more the
techniques introduced hereabove but, at least in the Gaussian case, it is
easy to obtain
exact expressions through the use of auxiliary integrations, over both
commuting and anti-commuting variables.

We first rederive our previous results for positive moments (i.e.
$\epsilon_l =+1$ for all $l$'s) . Let us introduce $M$ Grassmann variables
$\bar{c_a},c_a$ and an integration normalized to
\be \int \frac{d\bar{c}dc}{\pi} \bar{c} c = 1 . \ee
Then, for an hermitian $M\times M$ matrix $X$, one has
\be \label {Grass} \det(\la -X) = N^{-M}\int \prod_1^M
\frac{d\bar{c_a}dc_a}{i\pi} \exp{iN\sum_{a,b}[\bar{c_a}(\la \delta_{a,b}
-X_{a,b})c_b]}.\ee
A product $\displaystyle \prod_1^K\det(\la_l-X)$ is represented  by a
product of $K$ integrals of the type (\ref{Grass}). At the end the random
matrix
$X$ appears in an expression of the form
\be  \exp-{iN\sum_{l=1}^K \sum_{a,b=1}^M X_{ab} {\bar c_a^{(l)}}c_b^{(l)}}.\ee
With the Gaussian probability weight (\ref{GAUSS}) we have
\be <\exp{iN\Tr AX}> = \exp {-\frac{N}{2}\Tr A^2}, \ee
and thus
\be <\prod_1^K \det (\la_l-X)> =  N^{-M}\int \prod_1^M
\frac{d\bar{c_a}dc_a}{\pi}
\exp{(N\sum_{l=1}^K i\la_l\g_{ll}+\frac{N}{2}\sum_{l,m=1}^K
\g_{lm}\g_{ml})}  \ee
with
\be \g_{lm} = \sum_{a=1}^N {\bar c_a^{(l)}} c_a^{(m)}.\ee
We can use an auxiliary $K\times K$ hermitian matrix $B$ to replace the
quadratic terms in $\g$ by
\be \exp {\frac{N}{2}\Tr \g^2} = (\frac{N}{2\pi})^{K^{2}/2}\int d^{K^2}B
\exp {( N \Tr\g B -\frac{N}{2}\Tr B^2)} .\ee
We are left with an integral over the Grassmannian variables
\ba &&N^{-MK}\int \prod_{a=1}^M\prod_{l=1}^K
\frac{d\bar{c_a}^{(l)}dc_a^{ (l)}}{i\pi}
\exp{N\sum_{l,m=1}^K(i\la_l\delta_{lm}+B_{lm} )
\sum_{a=1}^M \bar{c_a}^{(l)}c_a^{ (m)}}\nonumber \\
&&= \left(\det_{1\leq l,m\leq K} (\la_l\delta_{lm} - iB_{lm})\right)^M. \ea

We end up with an integral over a $K\times K$ hermitian matrix $B$ :
\be <\prod_1^K \det(\la_l-X)> =   (\frac{N}{2\pi})^{K^{2}/2}\int d^{K^2}B
\left\{ \det(\la_l\delta_{lm} - iB_{lm})\right\}^M
\exp { -\frac{N}{2}\Tr B^2}.\ee
Therefore, from this method as well, we have reduced the correlations of
the characteristic functions of the matrix, to an integral over $K^2$
variables.
If one is interested in the moments, i.e. $\la_l =\la$ for all $l$'s, one
may  take as variables the eigenvalues
$b_l$ of $B$ (which yields a factor $\D^2(\la_1,\cdots,\la_K)$), and
recover the previous expressions. For the $\la_l$'s non-equal, one must
first shift
the matrix $B$ of the diagonal matrix $i\la_l\delta_{lm}$, and then
integrate out the unitary group $SU(K)$ by the Itzykson-Zuber
formula \cite{Harish-Chandra,Itzykson,Duistermaat}, to reduce it, as
before, to an
integral over $K$ variables (a slightly different integral, but which may
be handled in the
large $N$-limit in an identical fashion).

In case  of negative moments the method is identical, except that we need
now ordinary commuting variables, instead of Grassmannian.
Indeed starting from
\be \frac{1}{\det (\la -X\pm i\e)} = N^{M}\int \prod_1^M \frac{d\phi^{*}_a
d\phi_a}{\pm i\pi} \exp{\pm iN\sum_{a,b}[\phi^{*}_a(\la \delta_{a,b}
-X_{a,b}\pm i\e \delta_{a,b})\phi_b]},\ee
one can introduce, for each factor $ (\det(\la_l-X))^{\e_l}$ an integration
over $M$ complex variables $(\phi^{*}_a,\phi_a)$ if
$\e_l=-1$, or over $M$ complex Grassmannian variables $(\bar{c_a}, c_a)$ if
$ \e_l = +1$. The expectation value with the Gaussian weight $P(X)$ is then
immediate.
Of course for the negative moments, one must pay attention to the sign of
the infinitesimal imaginary part of the $\la$'s since there is a cut on the
real axis
along the support of Wigner's semi-circle.

Although the method is obvious and elementary, the notations can become
cumbersome and, rather than working out the most general case,
 and arbitrary choices for the
signs of the imaginary parts, we  restrict ourselves to an example.
 If we consider only negative powers,  we may follow identical
steps as hereabove  with positive powers, and we find
\ba <\prod_1^K\frac{1}{\det (\la_l-X+i\e)}> &=&
(\frac{N}{2\pi})^{\frac{K^{2}}{2}}\int d^{K^2}B \left\{ \det(\la_l\delta_{lm}-
B_{lm}+i\e \delta_{lm})\right\}^{-M}\nonumber\\
&\times&
\exp { -\frac{N}{2}\Tr B^2}.\ea
When all the $\la$'s are equal the r.h.s. simplifies to an integral over
$K$ variables $ \displaystyle \int \prod_1^K db_l
\frac{\D^2(b_1,\cdots,b_K)}{\prod_1^K (\la - b_l+i\e)^{M}}
\exp { -(\frac{N}{2}\sum_1^K b_l^2)}$.
For the $\la_l$'s non equal, after a shift of the matrix $B$ and the
integration over $SU(K)$, one obtains
\ba\label{negative} <\prod_1^K\frac{1}{\det (\la_l-X+i\e)}> =
(\frac{N}{2\pi})^{K(K+1)/2}\exp{ -\frac{N}{2}\sum_1^K \la_l^2}\nonumber \\
\times\int \prod_1^K
\frac{db_l}{(b_l-i\epsilon)^M}
\frac{\Delta(b_1,\cdots,b_K)}{\Delta(\la_1,\cdots,\la_K)}
\exp { -N\sum_1^K(\frac{1}{2}b_l^2 + b_l\la_l)},\ea
from which one could repeat easily the analysis of  section 3.

\section{Discussion}

     We have discussed the universal expressions for the moments of the
     characteristic polynomials in a random matrix theory, where the ensembles
     belong to the unitary family $(\beta = 2)$.

      We have shown that these universalities are related to the universality
      of the kernel in the Dyson's short distance limit.
      Since the statistics of the zeros of the $\zeta$-function
       follows the universal behavior of Gaussian unitary ensemble
      (GUE) \cite{KatzS,Montgomery}, the power moment of the $\zeta$-function
       also has to follow the universal behavior of GUE. We have studied
       here the characteristic polynomial, which corresponds to the
       $\zeta$-function on the critical line, and we have found a universal
       behavior for  the  moments of the characteristic polynomial.
      The universal number (\ref{M})
      appears indeed in the average of the moment of the $\zeta$-function,
      which was conjectured as (\ref{gamma}), $\gamma_K =
      \prod_{0}^{K-1} l!/(K + l)!$.

       Our method of the splitting the
       singularity by the introduction of the
       distinct $\la_i$ may be applied directly to the
       average of the power moment of the Riemann $\zeta$-function.
      We consider the average of the
       product of $\zeta (s_i)$, $s_i = \frac{1}{2} \pm i (\la_i +  t) $,
       \be
       F = \frac{1}{T} \int_0^T \prod_{i=1}^{2 K} \zeta( s_i) dt
       \ee
       where we choose $K$ positive $\la_i$'s and $K$ negative ones.
       If, at the end of the calculation, we set all the positive $\la_i$'s
equal to $\la$ and the negative ones to $-\la$,
one recovers the
       $2K$-th moment of the modulus  of the $\zeta$-function.
      When $T$ is large, the leading and the next leading terms of the
      derivative of $\ln F$ with respect to $\la_i$, are presumably given by
      \be\label{zetalogf}
      \frac{\partial \ln F}{\partial \la_i} \sim \pm i \ln T -
      \sum_{j\ne i} \frac{1}{\la_i - \la_j},
      \ee
      where the pole in the second term appears when
      two distinct $\la_i$ coincide,
      one of the $\la_i$'s with a plus sign and the other one
      with a minus sign. In the Appendix a discussion of the assumptions
leading to  (\ref{zetalogf}) is
      given.
       Then, after integration, we have, following a line of  arguments
similar to those of
      section 5,
      \be\label{zeta}
      F = \frac{c}{K!} \oint \prod_{i=1}^K \frac{du_i}{2 \pi i}
      \frac{\Delta^2(u)}{\prod_{i=1}^K\prod_{j=1}^{2K}
      (u_i - \la_j)} e^{- i \sum u_i \ln T}.
      \ee
      Therefore, if we let the $\la_i$'s coincide, we recover the integral
(\ref{G39}), which provides the universal coefficient $\gamma_K$.
 The coefficient $c$ is
      not determined by this method, which starts with the logarithmic
derivative of F, and an extra normalization condition
is needed.
     In (A.7) it will be argued that a coefficient $a_K$ is present in the
result, which is the residue
      at $s=1$ of a function $g_K(s)$ defined in the Appendix ; it is thus
plausible that the
      coefficient $c$ in (\ref{zeta}) is nothing but $c = a_K$.

      We have also investigated  negative moments as (\ref{negative}).
      This result may apply to the mean value of  negative moments
      of the $\zeta$-function. Indeed, the exponent $K^2$ of $\log T$
      for the negative integer $K$, has been conjectured
      \cite{Gonek}.

      For the symplectic and orthogonal case, $Sp(N)$ and $O(2N)$ ensembles,
      there may be also be a  correspondence between the random matrix results
      (\ref{spfinal2}), (\ref{onfinal2}) and the average values of the
      certain $L$-functions, with the same $\gamma_K$, as far as there is a
universality.
Existing conjectures \cite{CF} for the moment of the $L$
        function shows the same exponent
        $\frac{K}{2}{(K +1)}$ and $\frac{K}{2}(K - 1)$ for the symplectic and
        the orthogonal cases, however, the conjectured values of
        $\gamma_K$ is different from our result (\ref{spfinal2})
        for the symplectic case.

\begin{center}
{\bf Acknowledgement}
\end{center}
 This work was supported by the CREST of JST, and one of us (E.B.) is happy
to thank the organizers of
the third CREST meeting for the invitation extended to him.

\vskip 8mm
\setcounter{equation}{0}
\renewcommand{\theequation}{A.\arabic
{equation}}
{\bf Appendix : {Summation formula for the Riemann zeta-function}}
\vskip 5mm

  The Riemann $\zeta$-function is given by
  \be\label{A.1}
    \zeta(s) = \sum_{n=1}^{\infty} \frac{1}{n^s} = \prod_p ( \frac{1}{1 -
    \frac{1}{p^s}}).
  \ee
  where $p$ is a prime number. The $K$-th power of this function is written as
  \ba\label{A.2}
  [ \zeta(s) ]^K  &=& \sum_{n=1}^\infty \frac{d_K(n)}{n^s}\nonumber\\
   &=& \prod_p ( 1 + \frac{d_K(p)}{p^s} + \frac{d_K(p^2)}{p^{2s}} + \cdots ),
   \ea
   where $d_K(n)$ is the $K$-th Dirichlet coefficient. When $n$ is a power of
   the prime number,
    $d_K(p^j) = \Gamma(K + j)/\Gamma(K)j!$,
   (this follows easily from the definition of the Dirichlet coefficient
   $\displaystyle d_K(n) = \sum_{n_1 \cdots n_K = n} 1$).

  We consider now  the average of (\ref{A.2}) on the critical line $s =
\frac{1}{2} + i t$
  over a large interval  $T$,
  \be\label{A.3}
    \frac{1}{T} \int_0^T dt |\zeta( \frac{1}{2} + i t )|^{2 K} =
    \frac{1}{T} \int_0^T dt |\sum_{n=1}^\infty
    \frac{d_K(n)}{n^{\frac{1}{2} + i t}} |^2.
    \ee
 Expanding the sum $\displaystyle |\sum_{n=1}^{\infty}
    \frac{d_K(n)}{n^{s}} |^2 $, which appears in (\ref{A.3}), we  first
examine the diagonal terms ,
  \ba\label{A.4}
   \sum_{n=1}^\infty \frac{d_K^2(n)}{n^s} &=&
     \prod_p ( 1 + \frac{d_K^2(p)}{p^s} + \frac{d_K^2(p^2)}{p^{2s}} + \cdots)
     \nonumber\\
     &=& \prod ( 1 - p^{-s} )^{- K^2} ( 1 - \frac{K^2 ( K - 1)^2}{4} p^{-
2s} +
     \cdots )
     \nonumber\\
     &=& [\zeta(s)]^{K^2} g_K(s),
     \ea
     where
     \be
     g_K(s) = \prod_p [ ( 1 - p^{-s})^{K^2} \sum_{j=0}^\infty
     \frac{d_K^2(p^j)}{p^{j s}}].
     \ee
     The function $g_K(s)$ is an analytic function of $s$, including the
point $s=1$.

Let us examine the contribution of these diagonal terms given by
(\ref{A.4}) to (\ref{A.3}). Their contribution is conveniently found, if we
   apply the following inversion formula (Perron formula).
   \ba
      B(s) &=& \sum_{n=1}^\infty b_n n^{-s}\nonumber\\
      f(x) &=& \sum_{n\le x} b_n
      \ea
      Then, we have
      \be
      f(x) = \frac{1}{2 \pi i} \int_{c - i\infty}^{c + i\infty}
      B(s) x^s s^{-1} ds,
      \ee
in which $c$ is some arbitrary real positive number.
     Substituting  $b_n = d_K^2(n)$, and $B(s) = \zeta^{K^2}(s)
     g_K(s)$,
     we obtain, from the residue of the singularity  at $s = 1$,
     \be
     \sum_{n\le x} d_K^2(n) = \frac{g_K(1)}{\Gamma(K^2)} x \log^{K^2 - 1} x
+ O(x \log^{K^2 - 2} 3 x).
     \ee
     By a partial summation,  this approximate calculation yields ,
     \be
     \sum_{n\le T} \frac{d_K^2(n)}{n} \sim \frac{a_K}{\Gamma(K^2 + 1)}
      \log^{K^2} T
     \ee
     where $a_K = g_K(1)$.

    From these formulae, it is seen that the contribution of the diagonal
terms  to the average of the
    $K$-th power moment of the $\zeta$-function does take the asymptotic
form of
    (\ref{conj}). However, neglecting the off-diagonal terms, we failed to
reproduce the proper   coefficient $\gamma_K$,
    whose understanding clearly requires the off-diagonal products in
(\ref{A.4})
as well.

    A lower bound for the $2 K$-th moment is known \cite{Ramachandra}
    \be
    \int_{T-Y}^{T+Y} |\zeta(\frac{1}{2} + i t)|^{2K} dt \gg Y \log^{K^2}Y
    \ee
     where $\log^{\epsilon}T \le Y \le T$.
      An upper bound seems difficult to obtain, and (\ref{conj}) remains as
a conjecture, except for the $K=1 $ and $K = 2$ cases,for which
        it has been  derived.

     We note here the results and the conjecture of Montgomery
\cite{Montgomery}
     about the density of the zeros of Riemann $\zeta$-function and their
     correlation. When $\gamma$ is a zero on the critical line,
     $\zeta( \frac{1}{2}
     + i \gamma ) = 0$,
     \be\label{A.11}
        \sum_{0<\gamma \le T} 1 \ge (\frac{2}{3} + o(1)) \frac{T}{2 \pi} \log T
        \ee
        \be\label{A.12}
        \sum_{0<\gamma,\gamma'\le T, \alpha / L \le \gamma - \gamma'
        \le  \beta / L
        } 1 = ( 1 + o(1))[ \int_{\alpha}^{\beta} ( 1 - ( \frac{\sin \pi u}{
        \pi u})^2 ) du + \delta(\alpha,\beta)] T L
        \ee
        where $L = \log T/(2 \pi)$, and $\delta(\alpha,\beta) = 1$ for $0
        \in [\alpha,\beta]$, and  otherwise zero. Then (\ref{A.11}) is
equivalent to
        the average density of state in (\ref{conj}), with for  $K=1$,
$\gamma_K
        = a_K = 1$ and (\ref{A.12})
        is  equivalent to
        the pair correlation function in  random matrix theory.

        Les us present  the arguments which lead to the conjectured formula
(\ref{zetalogf}) ; we first assume
        assume  that $\la_1 - \la_2 \sim O((\ln T)^{-1})$ for  large $T$.
        The diagonal approximation for the product of $\zeta(s_1)$ and
        $\zeta(s_2)$, which earlier gave the expected behaviour for the
moment, but with a wrong coefficient, may thus
be applied here again, since we are taking a logarithmic derivative, which
is unsensitive to overall normalizations. Within this
assumption,  we obtain
        \ba
          \frac{\partial}{\partial \la_1} \log F &=&
          \frac{\partial }{\partial \la_1}\ln [\frac{1}{T}\int_1^T dt
          (\sum_{n=1}^{\infty} \frac{1}{
          n^{\frac{1}{2} + i \la_1 + it}}) (\sum_{n=1}^{\infty}
\frac{1}{n^{\frac{1}{2}
          - i \la_2 - i t}}) \zeta(s_3) \cdots \zeta (s_{2K})] \nonumber\\
          &\sim&
          \frac{\partial}{\partial \la_1} \ln [(\sum_{n<T}\frac{1}{n^{1 +
          i(\la_1 - \la_2)}})\frac{1}{T}\int_1^T dt \zeta(s_3)\cdots
           \zeta(s_{2K})]\nonumber\\
           &\sim& \frac{\partial}{\partial \la_1} \ln [ \int_{T_0}^T
           dx \frac{1}{x^{1 + i (\la_1 - \la_2)}}]\nonumber\\
           &\sim& - i \ln T - \frac{1}{\la_1 - \la_2}
           \ea
           We have considered up to now what happens when $\la_1 - \la_2$
is small, but we should repeat the same arguments for
the Dyson limit in which all pairs $\la_1 - \la_j$  are of order $(\log
T)^{-1}$. Therefore, when  one sums over  all possible combinations, one
obtains
           (\ref{zetalogf}).

\end{document}